\documentclass{article}

\usepackage{arxiv}

\usepackage[utf8]{inputenc}
\usepackage[T1]{fontenc}
\usepackage[english]{babel}
\usepackage{amsmath}
\usepackage{amssymb}
\usepackage{amsfonts}
\usepackage{booktabs}
\usepackage{array}
\usepackage{graphicx}
\usepackage{xcolor}
\usepackage{microtype}
\usepackage{etoolbox}
\usepackage{ragged2e}
\usepackage{multicol}
\usepackage{algpseudocode}
\usepackage[numbers,sort&compress]{natbib}
\usepackage{url}
\usepackage{doi}
\usepackage{hyperref}
\usepackage{xr-hyper}

\graphicspath{{figure/}}
\AtBeginEnvironment{multicols}{\justifying}

\newcounter{algorithm}
\renewcommand{\thealgorithm}{\arabic{algorithm}}

\newcommand{\orcidicon}[1]{\href{https://orcid.org/#1}{\includegraphics[scale=0.06]{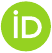}}}

\title{Dynamic sampling of non-stationary spontaneous activity in dissociated neuronal networks}

\author{%
\normalfont
\orcidicon{0009-0000-8487-3826}\hspace{1mm}Kazushi Takehana$^{1}$,
\orcidicon{0009-0009-6445-2587}\hspace{1mm}Dai Akita$^{1}$, and
\orcidicon{0000-0002-1834-3832}\hspace{1mm}Hirokazu Takahashi$^{1,2}$\thanks{Author to whom correspondence should be addressed.}\\
$^{1}$Graduate School of Information Science and Technology, The University of Tokyo, Tokyo, Japan\\
$^{2}$International Research Center for Neurointelligence (WPI-IRCN), The University of Tokyo, Tokyo, Japan\\
\texttt{k.takehana@ne.t.u-tokyo.ac.jp, d.akita@ne.t.u-tokyo.ac.jp, takahashi@i.u-tokyo.ac.jp}
}

\date{}

\renewcommand{\headeright}{Preprint}
\renewcommand{\undertitle}{Preprint}
\renewcommand{\shorttitle}{Dynamic sampling of spontaneous activity}

\hypersetup{
pdftitle={Dynamic sampling of non-stationary spontaneous activity in dissociated neuronal networks},
pdfsubject={q-bio.NC, q-bio.QM},
pdfauthor={Kazushi Takehana, Dai Akita, Hirokazu Takahashi},
pdfkeywords={high-density microelectrode array, adaptive electrode selection, non-stationary spontaneous activity, synchronized burst, Thompson sampling}
}

\begin{document}
\maketitle

\begin{abstract}
\textbf{Objective.}
To develop and evaluate an adaptive electrode-selection method for tracking non-stationary spontaneous activity during long-term high-density microelectrode array (HD-MEA) recordings under a fixed channel budget.

\textbf{Approach.}
We formulated electrode allocation as a sequential subset-selection problem and used a discounted Poisson--Gamma model with Thompson sampling. The method updated electrode-specific activity estimates from observed spike counts and reallocated a fixed channel budget over time. We evaluated it by offline replay of nine 34 h HD-MEA recordings, selecting 100 electrodes from 529 densely routed candidates, and in a representative online recording using 1,024 routed electrodes.

\textbf{Main results.}
Across offline recordings, the top-100 active-electrode set changed substantially, reaching 47.8\% turnover at 34 h. The Bayesian method captured the largest fraction of the spikes available to an oracle selector among the tested strategies and exceeded static selection by 17.2 percentage points at the final time point. In the online recording, adaptive selection captured the first synchronized burst and supported center-of-activity trajectory analysis.

\textbf{Significance.}
Uncertainty-aware exploration and temporal discounting can improve HD-MEA recording efficiency under fixed readout constraints, providing a basis for adaptive sensing of evolving neural activity.
\end{abstract}


\section{Introduction}

\begin{figure}
    \centering
           \IfFileExists{figure/fig1.pdf}{%
           \includegraphics[width=0.92\textwidth]{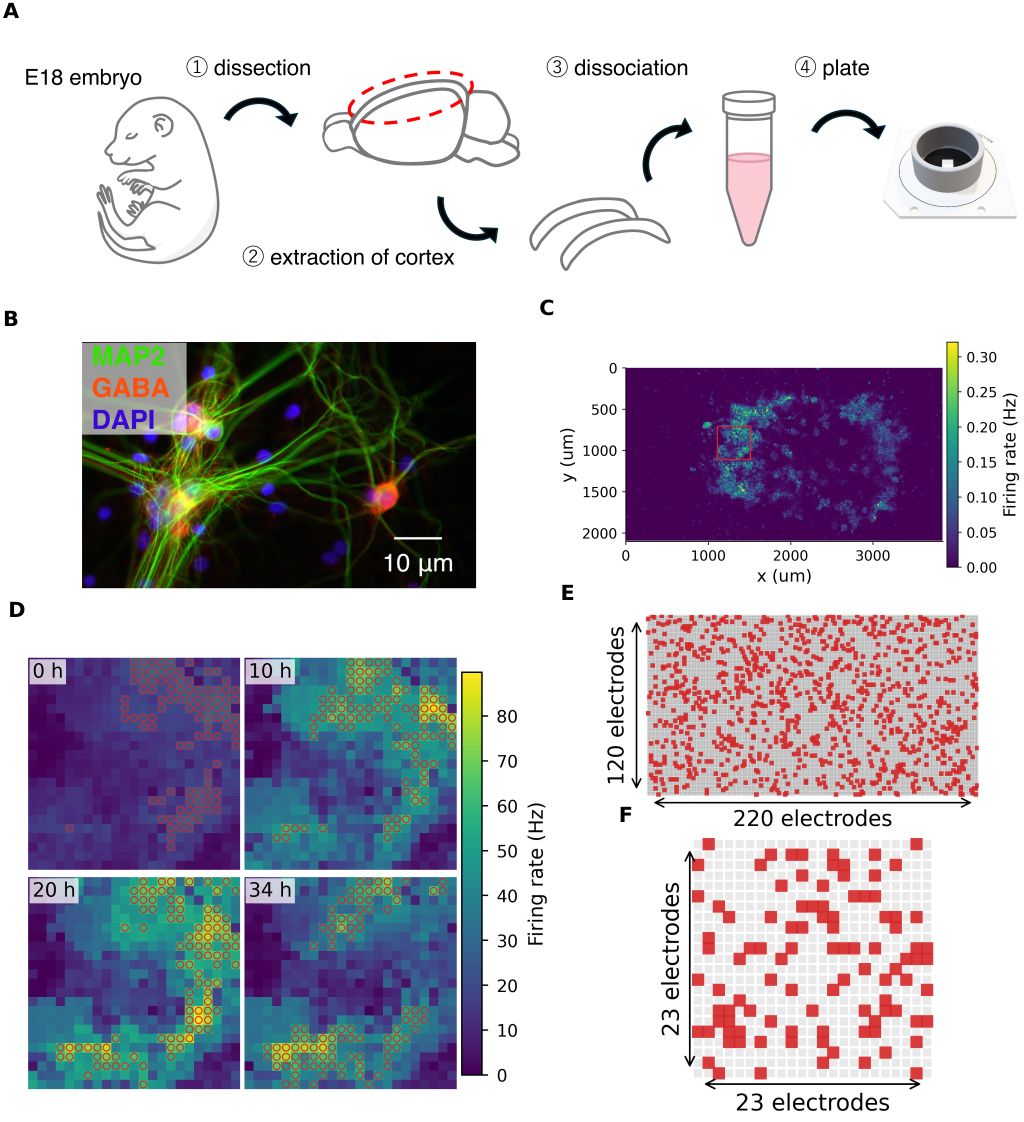}}{%
           }%
       \caption{
       Experimental setup and offline evaluation framework.
       \textbf{(A)} Culture preparation workflow. Cortices were isolated from embryonic day 18 (E18) Wistar rats, dissociated using trypsin, and plated onto HD-MEAs coated with polyethylenimine (PEI) and laminin.
       \textbf{(B)} Representative fluorescence image of a dissociated neuronal culture at day in vitro 15 (DIV15), immunostained for microtubule-associated protein 2 (MAP2; green) and gamma-aminobutyric acid (GABA; red), with nuclei counterstained with 4',6-diamidino-2-phenylindole (DAPI; blue).
       \textbf{(C)} Full-array firing-rate map constructed from an initial scan of the neuronal culture. To cover the full 26,400-electrode array, 1,024 electrodes were recorded for 60 s per configuration across 26 configurations. The recording area was 3.85 $\times$ 2.10 mm$^2$, with 17.5 $\mu$m electrode pitch and 11.5 $\times$ 11.5 $\mu$m$^2$ electrode size. The red square indicates the local 23 $\times$ 23 electrode region used for dense long-term recording.
       \textbf{(D)} Long-term activity maps recorded from the 529 routed electrodes shown in (C) at 0, 10, 20, and 34 h after recording onset. Red open circles indicate the 100 electrodes with the highest firing rates at each time point.
       \textbf{(E)} Schematic illustration of selecting 1,024 electrodes from the full 26,400-electrode array. Red squares indicate selected electrodes.
       \textbf{(F)} Schematic illustration of the offline evaluation setting, in which 100 electrodes were selected from the 529-electrode candidate set. Red squares indicate selected electrodes.
       }
       \label{fig1}
       
\end{figure}

\begin{multicols}{2}
Planar microelectrode arrays (MEAs) were developed to monitor bioelectric activity from cultured cells using multiple substrate-integrated electrodes \cite{thomas1972miniature}. Passive MEAs with tens of recording sites, including widely used 60- or 64-electrode formats, then became a standard platform for in vitro extracellular electrophysiology; early planar-array studies established that cultured neurons could be electrically stimulated and recorded from on the same substrate \cite{jimbo1992electrical,obien2015revealing}. Because they are compatible with repeated extracellular measurements from the same preparation, MEAs are particularly useful for dissociated neuronal cultures, which are now widely used to study how collective neural dynamics emerge, reorganize, and stabilize during development \cite{van2004long,chiappalone2006dissociated,wagenaar2006extremely,tetzlaff2010self,cabrera2021early,yada2017development}. As synaptic connectivity matures, spontaneous activity typically shifts from sparse, weakly coordinated firing to synchronized bursting and structured population dynamics \cite{chiappalone2006dissociated,wagenaar2006extremely,wagenaar2006persistent,masquelier2013network,cabrera2021early,yada2017development}. Because these dynamics are closely linked to network development, plasticity, prediction-like responses, and state transitions, long-term electrophysiological measurements are central to understanding how cultured neuronal networks self-organize over time \cite{yaron2025dissociated,zhang2025deviance,ikeda2025emergent}.

High-density microelectrode arrays (HD-MEAs) have expanded these measurements by combining dense spatial sampling with the temporal resolution needed to resolve extracellular spikes, burst propagation, and subcellular electrical footprints \cite{frey2010switch,bakkum2013tracking,ballini20141024,muller2015high,obien2015revealing,obien2019large}. A key step was the integration of CMOS circuitry beneath the electrode array. In switch-matrix-based designs, introduced to make dense electrode grids experimentally usable, many electrodes can be connected through programmable routing to a smaller set of recording and stimulation channels \cite{frey2010switch,ballini20141024,muller2015high}. This architecture makes it possible to examine activity across multiple scales, from local cellular signals to culture-wide network events. The spatial resolution is particularly valuable in dissociated cultures, where spontaneous activity can include localized initiation sites, propagating bursts, and recurring spatiotemporal motifs \cite{wagenaar2006persistent,yada2016state,tajima2017locally}.

The cost of this scaling is a readout bottleneck. In many switch-matrix CMOS HD-MEAs, the electrode count greatly exceeds the number of low-noise amplifiers, analog-to-digital converters, and data paths available for simultaneous recording, so only a limited subset of electrodes can be routed to recording channels at any given time \cite{PerezPrieto2021}. This constraint reflects the separation between dense electrode sites and a smaller number of low-noise readout chains: electrode pixels mainly implement routing switches and memory elements to preserve spatial resolution, whereas amplifiers, filters, and analog-to-digital converters are placed outside the sensing area and shared by user-selected electrodes \cite{ballini20141024}. This architecture helps maintain high spatial resolution, low noise, and low power, but it makes simultaneous recording a routing problem from many electrodes to fewer channels. In the platform used here, for example, 26,400 electrodes are available on the array, but at most 1,024 electrodes can be routed simultaneously \cite{ballini20141024}. Other CMOS architectures, including active-pixel and very-large-scale arrays, have demonstrated more parallel or even full-array readout when the sensing electronics and data pipeline are designed for that purpose \cite{berdondini2009active,tsai2017very}. Recent large-area CMOS-MEA technology has extended this direction further, enabling field-potential imaging with 236,880 electrodes over a millimeter-scale area at single-neuron resolution \cite{suzuki2023largearea}. Such systems reduce the hardware need to choose among electrodes, but they also highlight a complementary challenge: as recording scales grow, data volume, storage, and downstream analysis can still make task-dependent subset selection useful. For widely used switch-matrix HD-MEAs, electrode selection is an immediate experimental constraint. In practice, this limitation is often handled by first scanning the array to identify active sites and then assigning the available channels to a fixed electrode subset for subsequent recording. Such scan-and-commit workflows are practical, but they assume that informative electrodes remain informative over the measurement period.

For spontaneous activity in dissociated neuronal networks, this assumption is fragile. Firing rates, burst statistics, synchrony, and propagation patterns change over development \cite{chiappalone2006dissociated,wagenaar2006extremely,masquelier2013network,cabrera2021early,yada2017development}, and synchronized bursts can follow state-dependent propagation patterns rather than a single fixed origin \cite{yada2016state}. Long-term HD-MEA studies have also shown that changes in culture morphology and network configuration can be reflected in the recorded activity maps \cite{okawa2015chronic,Habibey2022}. Consequently, electrodes that are highly informative at one time point may become less informative later, even in the same culture. A static assignment can therefore miss activity that emerges away from the initially selected sites.

These changes span time scales that are directly relevant to electrode allocation. Over seconds to minutes, individual bursts can vary in recruited sites and propagation structure \cite{van2004long,wagenaar2006extremely,yada2016state,tajima2017locally,rolston2007precisely}; over hours, burst-related activity can exhibit stable probabilistic structure that requires spatial coverage beyond instantaneous firing rate \cite{van2004long,wagenaar2006persistent}; and across many hours to days, activity maps can shift with development, morphology, and network configuration \cite{van2004long,chiappalone2006dissociated,wagenaar2006extremely,yada2017development,okawa2015chronic,Habibey2022}. For electrode selection, the useful subset is therefore a time-dependent representation of the current network state rather than a fixed anatomical label. Without updates, changes in burst recruitment or active-region location can reduce both spike yield and downstream spatial readouts such as propagation maps or center-of-activity trajectories \cite{chao2007region,yada2016state}. Thus, adaptive allocation is important for maintaining access to the evolving collective dynamics that make long-term cultures informative.

The gap between the spatial richness of HD-MEAs and the limited number of simultaneously recordable channels raises a simple question: under a fixed channel budget, which electrodes should be selected next? Multiplexed neural recording hardware makes this trade-off explicit \cite{PerezPrieto2021}, and adaptive electrode selection has been explored to improve signal quality or neuron yield in other recording settings \cite{choi2020optimal}. More broadly, closed-loop in vitro electrophysiology has used ongoing activity to guide stimulation, including control of bursting in cortical cultures \cite{wagenaar2005controlling} and algorithmically selected training signals in cortical organoids \cite{robbins2026goal}. Although such work addresses adaptive stimulation rather than adaptive sensing, it illustrates a shift toward experiments in which the next measurement or intervention is chosen from ongoing neural activity. Adaptive electrode selection has been less examined for long-term spontaneous recordings, where the objective is to track a non-stationary collective process rather than only to maximize unit isolation.

This setting is naturally framed as online decision-making under uncertainty. Multi-armed bandit methods formalize the exploration--exploitation trade-off, and Bayesian approaches such as Thompson sampling propagate uncertainty across decisions and can be adapted to non-stationary environments \cite{auer2002finite,chapelle2011empirical,daniel2018tutorial,garivier2011upper,qi2025thompson,lattimore2020bandit}. From this perspective, HD-MEA electrode allocation is an adaptive sensing problem: each recording window provides partial information that guides the next channel assignment.

In this study, we formulate electrode selection for long-term spontaneous HD-MEA recordings as a sequential subset selection problem under non-stationarity. The experimental preparation, recording configuration, and downscaled evaluation setting are summarized in figure~\ref{fig1}. We propose a Bayesian adaptive framework that updates each electrode's expected usefulness from observed spike activity and selects the next subset under a fixed channel budget. By accounting for temporal variability, the method aims to improve recording efficiency relative to static or heuristic strategies and to connect the spatial information available on modern HD-MEAs with the finite readout resources imposed by current hardware.
\end{multicols}

\section{Method}

\begin{figure}
 \centering
        \IfFileExists{figure/fig2.pdf}{%
        \includegraphics[width=\textwidth]{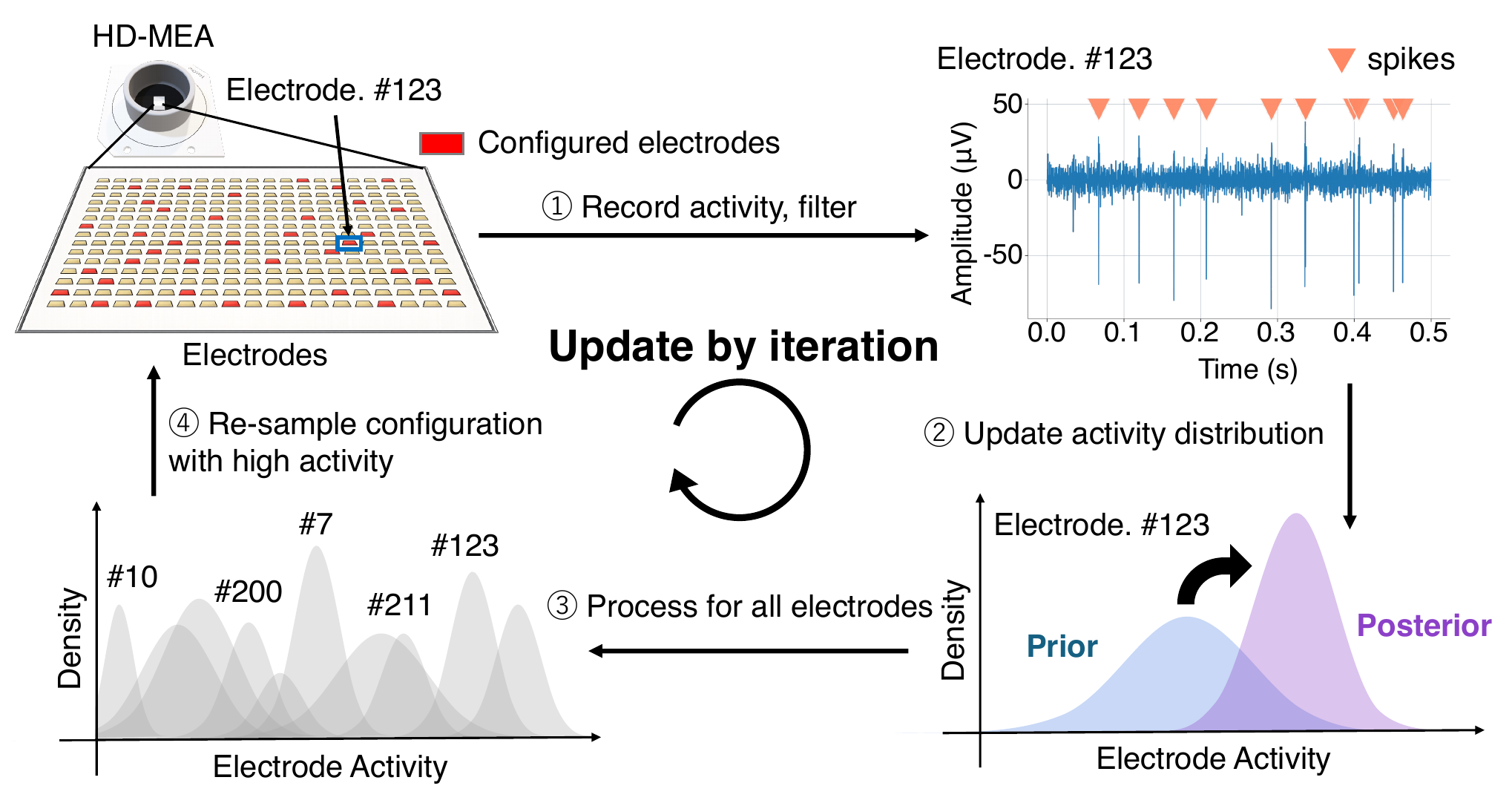}}{%
        }%
 \caption{Overview of the adaptive electrode selection framework. Red electrodes indicate the subset currently routed for recording under the fixed channel budget. In each recording window, extracellular traces from the routed electrodes are filtered and converted to spike counts. These observations update the probabilistic activity model for each observed electrode, shifting the posterior activity distribution from the prior according to the recorded spikes. The same update-and-selection procedure is repeated across electrodes and time windows. At the next step, the algorithm uses the updated activity distributions to choose a new electrode configuration, favoring electrodes with high estimated activity while retaining uncertainty-driven exploration.}
\label{fig2}

\end{figure}

\begin{table}[!t]
\centering
\refstepcounter{algorithm}\label{alg:discounted_pg_ts}
\begin{minipage}{0.97\textwidth}
\hrule
\vspace{0.4em}
\noindent\textbf{Algorithm \thealgorithm.} Discounted Poisson--Gamma Thompson sampling for electrode selection
\vspace{0.4em}
\hrule
\vspace{0.6em}
\small
\begin{algorithmic}[1]
\Statex \textbf{Input:} Candidate electrodes $\mathcal{C}$; channel budget $k$; window duration $\Delta$
\Statex \hspace{\algorithmicindent} Prior parameters $\alpha_0,\beta_0$; discount factor $\rho$; temperature $\tau$
\Statex \textbf{Output:} Selected electrode sets $S_t$; offline replay attainment scores
\State $\alpha_c \gets \alpha_0$, $\beta_c \gets \beta_0$ for all $c \in \mathcal{C}$
\For{each recording window $t=1,\ldots,T$}
    \State $\alpha_c \gets \rho \alpha_c$, $\beta_c \gets \rho \beta_c$ for all $c \in \mathcal{C}$
    \State Sample $\tilde{\lambda}_c \sim \mathrm{Gamma}(\alpha_c/\tau,\beta_c/\tau)$ for all $c \in \mathcal{C}$
    \State $S_t \gets \mathrm{TopK}(\{\tilde{\lambda}_c\}_{c \in \mathcal{C}}, k)$
    \State Record spike counts $y_{c,t}$ for all $c \in S_t$
    \State $\alpha_c \gets \alpha_c + y_{c,t}$, $\beta_c \gets \beta_c + \Delta$ for all $c \in S_t$
    \If{offline reference data are available}
        \State $R_t \gets \sum_{c \in S_t} y_{c,t}$
        \State $S_t^\ast \gets \mathrm{TopK}(\{y_{c,t}\}_{c \in \mathcal{C}}, k)$
        \State $R_t^\ast \gets \sum_{c \in S_t^\ast} y_{c,t}$
        \State $\mathrm{Attainment}_t \gets 100 R_t/R_t^\ast$
    \EndIf
\EndFor
\end{algorithmic}
\vspace{0.4em}
\hrule
\end{minipage}
\end{table}

\begin{table}[!t]
\centering
\refstepcounter{table}\label{tab:algorithm_parameters}
\begin{minipage}{\textwidth}
\centering
{\normalsize\bfseries Table \thetable.} Parameters used for offline replay and the representative online recording. The offline replay used $k_{\mathrm{off}}=100$ selected electrodes from 529 candidates. The online recording used the hardware channel budget of $k_{\mathrm{on}}=1024$ routed electrodes from the full 26,400-electrode array; the same Bayesian update interval and hyperparameters were used.
\vspace{0.8em}

\small
\begin{tabular*}{\textwidth}{@{\extracolsep{\fill}}llll@{}}
\hline
Parameter & Symbol & Value & Used in \\
\hline
Candidate electrodes, offline replay & $|\mathcal{C}_{\mathrm{off}}|$ & 529 & Offline replay \\
Selected electrodes, offline replay & $k_{\mathrm{off}}$ & 100 & Offline replay \\
Candidate electrodes, online recording & $|\mathcal{C}_{\mathrm{on}}|$ & 26,400 & Online recording \\
Selected electrodes, online recording & $k_{\mathrm{on}}$ & 1,024 & Online recording \\
Bin width / update interval & $\Delta$ & 1800 s (30 min) & Offline replay and online recording \\
Seed runs & $N_{\mathrm{seed}}$ & 50 & Offline randomized methods \\
Exploration rate & $\varepsilon$ & 0.2 & Offline $\varepsilon$-greedy \\
Decay factor & $d$ & 1.0 & Offline $\varepsilon$-greedy \\
Initial exploration bins & -- & 1 & Offline $\varepsilon$-greedy \\
Prior shape & $\alpha_0$ & 5.0 & Proposed method, offline and online \\
Prior rate & $\beta_0$ & 1.0 s & Proposed method, offline and online \\
Discount factor & $\rho$ & 0.7 & Proposed method, offline and online \\
Temperature & $\tau$ & 1.0 & Proposed method, offline and online \\
\hline
\end{tabular*}
\end{minipage}
\end{table}

\begin{multicols}{2}
\subsection{Culture preparation}

All animal procedures conformed to the Guiding Principles for the Care and Use of Animals in the Field of Physiological Science published by the Japanese Physiological Society. The experimental protocol was approved by the Committee on the Ethics of Animal Experiments at the Graduate School of Information Science and Technology, the University of Tokyo (A2024IST003).

HD-MEA chips were first cleaned with 1 mL of 1\% Tergazyme (Sigma-Aldrich, St Louis, MO, USA) for 2 h at room temperature. After removing the detergent, the chips were rinsed three times with sterilized water, immersed in ethanol for 30 min, and rinsed three additional times. The electrode area was then covered with 1 mL of pre-warmed plating medium (Neurobasal Plus, Thermo Fisher Scientific, Waltham, MA, USA), protected from drying, and kept in an incubator for at least 2 days before coating.

Before cell seeding, the chips were rinsed three times with sterilized water. Polyethylenimine (Supelco, Bellefonte, PA, USA) was diluted to 0.07\% in borate buffer (Thermo Fisher Scientific), and 50 $\mu$L of the solution was applied to the electrode area. The chips were incubated overnight, washed three times, and coated with 5 $\mu$L of laminin (20 $\mu$g/mL; Sigma-Aldrich) for 1 h.

Pregnant Wistar rats were anesthetized with inhaled isoflurane (Viatris, Canonsburg, PA, USA) and euthanized by decapitation. After abdominal disinfection with ethanol, the uterus was removed and transferred to Hank's Balanced Salt Solution (Life Technologies, Carlsbad, CA, USA). Three embryonic day 18 fetuses were collected, and cortical tissue was dissected from the isolated brains.

The cortical tissue was incubated in 2 mL of 0.25\% Trypsin-EDTA (Thermo Fisher Scientific) for 20 min, with gentle agitation every 5 min. The enzymatic reaction was stopped by transferring the tissue to plating medium. The tissue was then moved to fresh medium and mechanically dissociated by trituration. The cell suspension was passed through a 40 $\mu$m cell strainer (Falcon, Corning, NY, USA), and plating medium was added to adjust the cell density to 38,000 cells per 5 $\mu$L.

After removing the laminin solution, 5 $\mu$L of the cell suspension was placed onto the electrode area of each chip. The chips were incubated for 120 min to allow cell attachment, after which 0.6 mL of plating medium was added. Cultures were maintained at 36.5$^\circ$C in 5\% CO$_2$. To reduce evaporation, each chip was covered with its lid and placed in a 90 mm dish together with a 35 mm dish containing sterilized water.

The culture-preparation workflow is summarized in figure~\ref{fig1}(A), and a representative DIV15 culture prepared on an HD-MEA is shown in figure~\ref{fig1}(B).

\subsection{Electrical long-term recording}

Spontaneous extracellular activity was recorded using a CMOS-based HD-MEA platform (MaxOne, MaxWell Biosystems, Zurich, Switzerland). The array contains 26,400 electrodes over a 3.85 $\times$ 2.10 mm$^2$ recording area. The electrode pitch is 17.5 $\mu$m, and each electrode is 11.5 $\times$ 11.5 $\mu$m$^2$. The full-array activity map and local dense-recording region are shown in figure~\ref{fig1}(C). Each densely routed reference recording used for offline replay lasted 34 h and provided dense activity measurements from a local electrode region. This duration was chosen because, under the approximately 1 mL medium volume used on the chip during recording, stable long-term measurements were reliably maintained up to about 34 h. Raw extracellular traces were band-pass filtered between 300 and 3000 Hz before spike detection, and detected spikes were converted to spike counts for each recording window. Recording metadata and additional candidate-region maps are provided in supplementary table~\ref{supp-tab:recording_metadata} and supplementary figure~\ref{supp-fig:s1_candidate_regions}.

Network bursts were detected from the cumulative spike train using the ISI$_N$ framework described by Bakkum \textit{et al} \cite{bakkum2014parameters}. We used the same setting throughout all burst analyses: $N=100$. Thus, intervals in which 100 consecutive population spikes occurred within 100~ms were treated as burst candidates for subsequent raster and centroid analyses. Additional spike-event and burst-detection details are provided in supplementary note~\ref{supp-suppnote:spike_burst_detection}.

The densely routed recordings were used as reference data for the offline evaluation of electrode-selection algorithms described below. Activity maps from recording onset, 10 h, 20 h, and 34 h after recording onset are shown in figure~\ref{fig1}(D), illustrating the temporal redistribution of highly active electrodes during the recording.

\subsection{Dynamic electrode-selection algorithm}

An overview of the proposed dynamic electrode-selection procedure is shown in figure~\ref{fig2}. In brief, the algorithm records activity from the currently configured electrodes, updates a probabilistic model of electrode activity from the observed spike counts, and uses the updated uncertainty to choose the next electrode configuration.

The physical electrode-selection problem on the HD-MEA is to choose up to 1024 routed electrodes from the full 26,400-electrode array, as illustrated in figure~\ref{fig1}(E). Direct evaluation of an adaptive policy in this setting is difficult because the activity on unselected electrodes is not observed and the oracle selection cannot be computed. We therefore evaluated the algorithms in a downscaled offline setting. A dense 23 $\times$ 23 region, corresponding to 529 electrodes, was routed over an active area and recorded continuously for 34 h. These 529 electrodes were treated as the candidate set, and each algorithm selected 100 electrodes at each decision step, corresponding to the offline evaluation setting in figure~\ref{fig1}(F). This setting preserves the central constraint of the original problem: only a subset of electrodes can be observed at each time, but the densely recorded reference data allow the oracle to be computed retrospectively. Detailed replay and baseline definitions are provided in supplementary note~\ref{supp-suppnote:offline_replay}.

Let $\mathcal{C}$ denote the candidate electrode set and let time be divided into recording windows indexed by $t=1,\ldots,T$. The duration of each window is denoted by $\Delta$, and the spike count of electrode $c \in \mathcal{C}$ in window $t$ is denoted by $y_{c,t}$. At each step, the algorithm chooses $k$ electrodes to record,
\begin{equation}
S_t \subset \mathcal{C}, \qquad |S_t|=k ,
\end{equation}
where $S_t$ is the selected electrode set and $k=100$ in the offline evaluation. Only the spike counts from these selected electrodes are made available to the algorithm. The captured activity in each window was defined as the total spike count on the selected electrodes,
\begin{equation}
R_t = \sum_{c \in S_t} y_{c,t}.
\label{eq:reward_def}
\end{equation}
For reporting in firing-rate units, we used $\mathrm{FR}_t=R_t/\Delta$.

Here, $\mathrm{TopK}(\{v_c\}_{c \in \mathcal{C}},k)$ denotes the $k$ electrodes with the largest values $v_c$.

For each recording window, we also computed an oracle subset from the full offline recording. This oracle simply selected the $k$ electrodes with the largest spike counts in that same window,
\begin{equation}
S_t^\ast =
\mathrm{TopK}\left(
\{y_{c,t}\}_{c \in \mathcal{C}}, k
\right),
\end{equation}
and the corresponding oracle activity was $R_t^\ast=\sum_{c \in S_t^\ast} y_{c,t}$. We evaluated each method by the percentage of this oracle activity that it captured,
\begin{equation}
\mathrm{Attainment}_t =
100 \times \frac{R_t}{R_t^\ast}.
\label{eq:attainment}
\end{equation}
Because firing rate is obtained by dividing spike count by the same window duration $\Delta$, the same percentage is obtained from firing rates.

To quantify non-stationarity of the activity maps, we computed the turnover of the most active electrodes in the densely recorded reference data. Let
\begin{equation}
A_t =
\mathrm{TopK}\left(
\{y_{c,t}\}_{c \in \mathcal{C}}, k
\right)
\end{equation}
denote the top-$k$ active electrodes in window $t$. The turnover relative to the initial active set was defined as
\begin{equation}
\mathrm{Turnover}_t =
100 \times
\left(
1-\frac{|A_1 \cap A_t|}{k}
\right).
\label{eq:turnover}
\end{equation}
This measure is 0\% when the same top-$k$ electrodes are retained and increases as initially active electrodes are replaced by newly active electrodes.

Algorithm~\ref{alg:discounted_pg_ts} summarizes the full proposed procedure. The attainment calculation is included only for offline evaluation, where the reference activity of all candidate electrodes is available.

We used captured spike count as the primary reward because it is directly observable from routed channels, independent of hardware-specific spike-sorting assumptions, and provides a simple proxy for preserving downstream spatial summaries such as center-of-activity trajectories. This choice also makes the offline oracle well defined: the best subset in each window is the subset that captures the largest number of spikes under the same channel budget.

We modeled the spike count of each electrode using a Poisson--Gamma model. For readability, the time subscript is omitted from the model parameters below; the parameters are updated after every recording window. The current firing rate of electrode $c$ is denoted by $\lambda_c$, and the expected spike count in one window is $\lambda_c\Delta$:
\begin{equation}
y_{c,t}
\sim \mathrm{Poisson}(\lambda_c\Delta).
\label{eq:poisson_model}
\end{equation}
The algorithm represents uncertainty about $\lambda_c$ with a Gamma distribution,
\begin{equation}
\lambda_c
\sim \mathrm{Gamma}(\alpha_c,\beta_c),
\label{eq:gamma_prior}
\end{equation}
using the shape--rate parameterization.

In the shape--rate parameterization, the current expected firing rate is
\begin{equation}
\mathrm{mean}(\lambda_c)=\frac{\alpha_c}{\beta_c}.
\end{equation}
The Gamma distribution was used because it can be updated directly from Poisson spike counts. All electrodes were initialized with the same prior parameters listed in table~\ref{tab:algorithm_parameters}.

To account for non-stationarity in the activity distribution, old observations were discounted before each update \cite{garivier2011upper,qi2025thompson}. At the beginning of each window, the sufficient statistics were scaled as
\begin{equation}
\alpha_c \leftarrow \rho \alpha_c, \qquad
\beta_c \leftarrow \rho \beta_c
\quad (c \in \mathcal{C}),
\label{eq:discount}
\end{equation}
where $0<\rho\leq1$ is the discount factor. This keeps the current mean $\alpha_c/\beta_c$ unchanged, but increases uncertainty when $\rho<1$, allowing the algorithm to revisit electrodes whose activity may have changed. After observing the selected electrodes in window $t$, only those electrodes were updated:
\begin{equation}
\alpha_c \leftarrow \alpha_c+y_{c,t}, \qquad
\beta_c \leftarrow \beta_c+\Delta
\quad c \in S_t .
\label{eq:gamma_update}
\end{equation}
Electrodes not selected in the current window were not assigned artificial observations; only the discounting step was applied to them.

Electrodes were selected by Thompson sampling \cite{lattimore2020bandit}. In each window, one plausible firing rate was sampled for every electrode from its current Gamma distribution. To tune the degree of exploration, we used a temperature parameter $\tau>0$:
\begin{equation}
\tilde{\lambda}_c
\sim
\mathrm{Gamma}\left(
\frac{\alpha_c}{\tau},
\frac{\beta_c}{\tau}
\right).
\label{eq:ts_sample}
\end{equation}
The algorithm then selected the $k$ electrodes with the largest sampled rates,
\begin{equation}
S_t =
\mathrm{TopK}\left(
\{\tilde{\lambda}_c\}_{c \in \mathcal{C}}, k
\right).
\label{eq:ts_topk}
\end{equation}
This rule usually selects electrodes with high estimated firing rates, while still giving uncertain electrodes a chance to be sampled. The temperature scaling preserves the mean $\alpha_c/\beta_c$; larger $\tau$ increases exploration, whereas smaller $\tau$ makes the policy more exploitative.

We compared the proposed method with static selection, random selection, and a combinatorial $\varepsilon$-greedy baseline. Static selection chose the 100 electrodes with the largest spike counts in the first 30~min window of the dense 529-electrode reference recording and kept this subset fixed throughout the offline replay. Random selection chose 100 electrodes uniformly at random in each window.

The combinatorial $\varepsilon$-greedy baseline maintained an online mean estimate $\hat{\mu}_c$ for each electrode. At window $t$, the number of randomly sampled electrodes, $k_{\mathrm{rand}}(t)$, was set to the nearest integer to $\varepsilon_t k$. The baseline selected these electrodes at random and filled the remaining $k-k_{\mathrm{rand}}(t)$ positions with the electrodes that had the largest current mean estimates. During the initial exploration period, all selected electrodes were sampled randomly. For each selected electrode, the online mean was updated as
\begin{equation}
\hat{\mu}_c \leftarrow
\hat{\mu}_c +
\frac{1}{n_c}
\left(y_{c,t}-\hat{\mu}_c\right),
\end{equation}
where $n_c$ is the number of times electrode $c$ has been selected, including the current observation. The exploration rate was exponentially decayed as $\varepsilon_{t+1}=\varepsilon_t d$. In the main comparison, $d=1.0$, so the $\varepsilon$-greedy exploration rate was held constant; decayed variants are evaluated in supplementary figure~\ref{supp-fig:hyperparameter_sensitivity}(D,E).

The parameter values used for the offline evaluation and representative online recording are summarized in table~\ref{tab:algorithm_parameters}. Because random selection, combinatorial $\varepsilon$-greedy selection, and Thompson sampling depend on stochastic choices, each stochastic method was run 50 times with different random seeds for each recording. For dataset $i$ and seed run $r$, a distinct random seed was used. The 50 seed runs were first averaged within each recording. Time-resolved traces and turnover values were then summarized across the nine recordings as the mean with a two-sided 95\% confidence interval, calculated at each time point as $\bar{x}\pm t_{0.975,n-1}s/\sqrt{n}$ with $n=9$ recordings; lower confidence limits were clipped at zero. Final attainment values for the proposed method and static selection were compared across recordings using a two-sided exact paired sign-flip randomization test. For each recording, after averaging over seed runs, we subtracted the final attainment of static selection from the final attainment of the proposed method. The test statistic was the absolute value of the mean of these paired differences, and the exact null distribution was obtained by enumerating all $2^9$ possible sign flips.
\end{multicols}

\section{Results}

\begin{figure}[t]
\centering
       \IfFileExists{figure/fig3.pdf}{%
       \includegraphics[width=\textwidth]{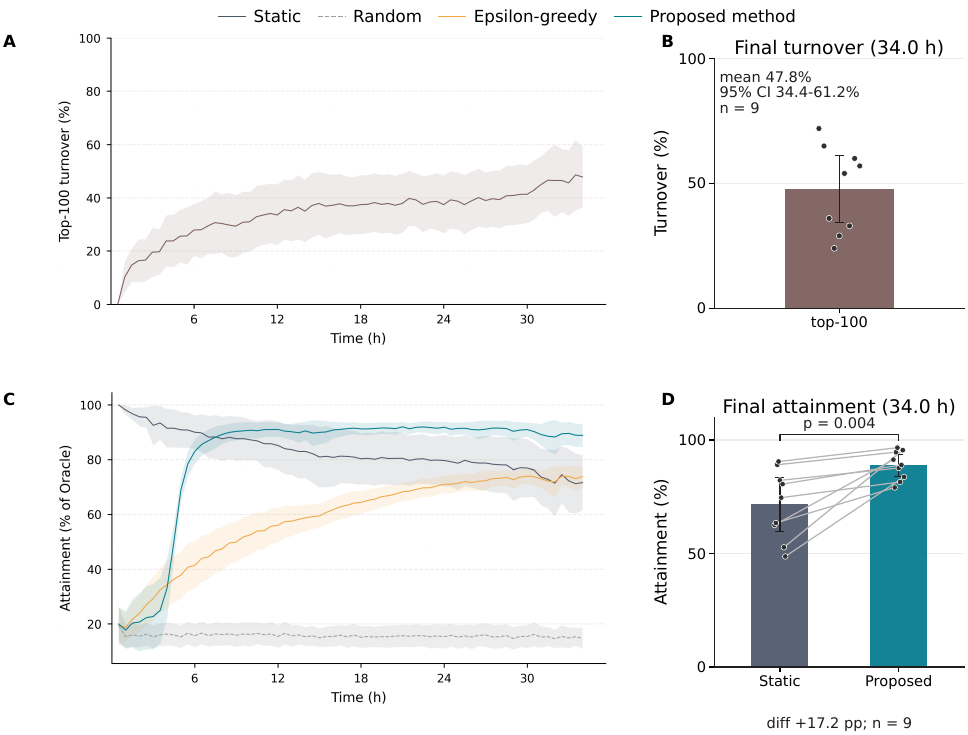}}{%
       }%
\caption{Non-stationarity of active electrodes and offline evaluation of adaptive electrode selection. \textbf{(A)} Turnover of the top-100 active electrodes over the 34 h recording, computed relative to the top-100 electrodes in the initial 30 min window. \textbf{(B)} Final top-100 turnover at 34 h across nine recordings. \textbf{(C)} Oracle-relative attainment for static selection, random selection, combinatorial $\varepsilon$-greedy selection, and the proposed method. Attainment is the percentage of oracle spike-count activity captured by the selected electrodes in each time window. \textbf{(D)} Final attainment at 34 h for static selection and the proposed method across nine recordings. Curves show the across-recording mean for each 30 min bin; stochastic policies were averaged over 50 random seeds within each recording before this across-recording summary. Shaded regions indicate across-recording 95\% confidence intervals where applicable; circles indicate individual recordings. The final comparison in (D) was performed across recordings using a two-sided exact paired sign-flip randomization test.}
\label{fig3}
\end{figure}

\begin{figure}[htbp]
\centering
       \IfFileExists{figure/fig4.pdf}{%
       \includegraphics[width=\textwidth]{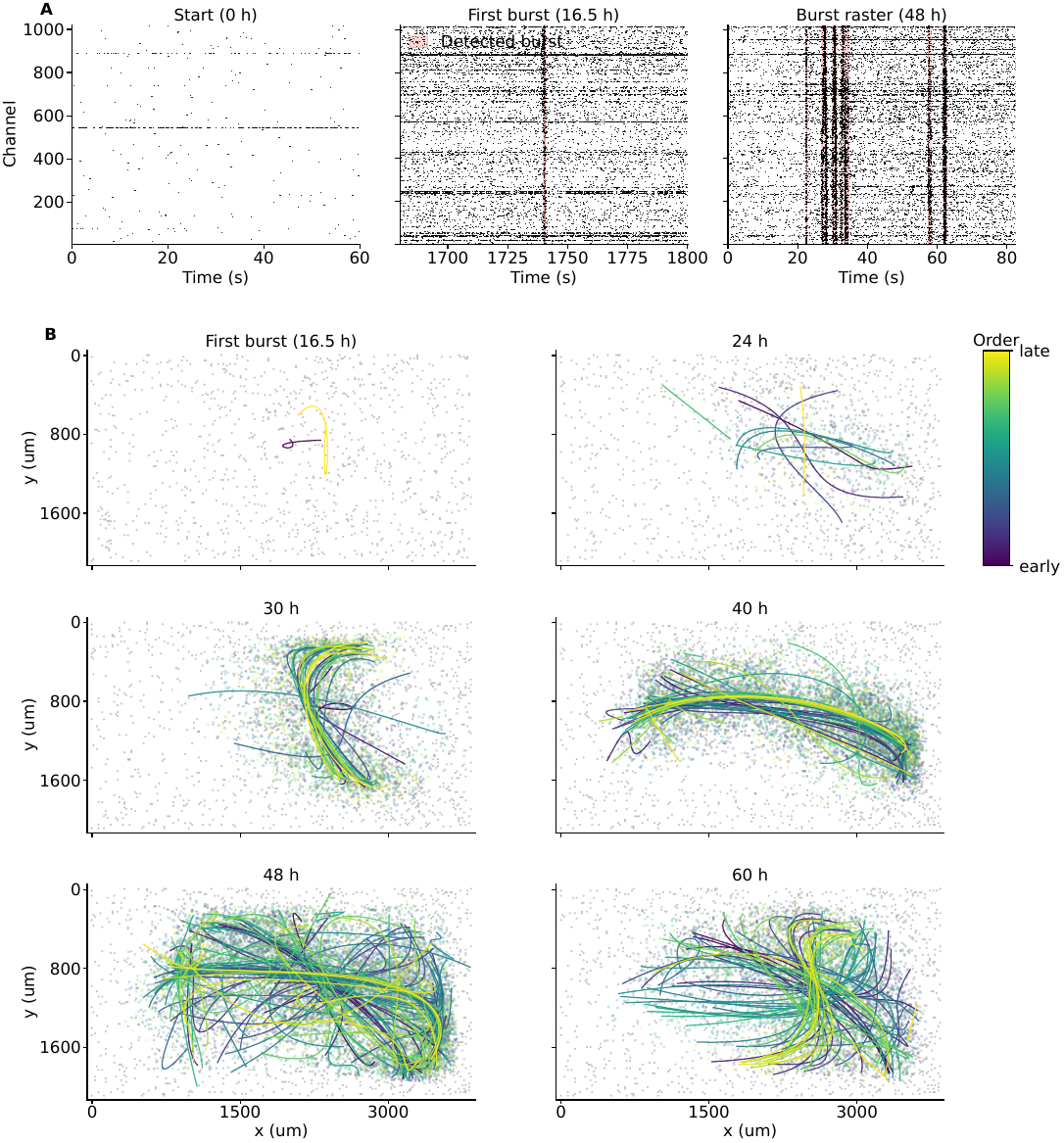}}{%
       }%
\caption{Representative online adaptive recording during network-burst emergence. Recording began 30~h after cell seeding, and the routed electrode set was updated every 30~min. The online recording used 1,024 routed electrodes; Bayesian hyperparameters other than the channel budget were taken from the best-performing offline setting in table~\ref{tab:algorithm_parameters}. Times indicate elapsed time from recording onset. \textbf{(A)} Raster plots show selected-electrode activity at onset, around the first detected network burst, and during later recurrent synchronized bursts. Activity was initially sparse, then appeared as localized events before developing into synchronized population bursts; shaded regions mark detected burst intervals. \textbf{(B)} Activity maps and trajectories show spike-count-weighted activity centroids within burst intervals. Spike events were assigned electrode coordinates, binned at 5~ms, and bins with fewer than 10 spikes were omitted; trajectories were smoothed with polynomial fits of degree at most three. Later trajectories became more recurrent, consistent with progressive organization of coordinated spatial activity.}
\label{fig4}
\end{figure}

\begin{multicols}{2}
\subsection{Spontaneous activity maps were non-stationary over long-term recording}

Before evaluating electrode-selection policies, we first quantified whether the active electrode set remained stable during the 34 h recordings. Using the densely recorded 529-electrode reference data, we identified the top-100 active electrodes in each 30 min window and measured their turnover relative to the initial top-100 set (figure~\ref{fig3}(A)).

The active set changed substantially over time. Turnover increased during the recording and reached 47.8\% at 34 h, with a 95\% confidence interval of 34.4--61.2\% across nine recordings (figure~\ref{fig3}(B)). Thus, by the end of the recording, roughly half of the electrodes that best represented the current activity were not part of the initially most active subset. This directly supports the central motivation for adaptive selection: a scan-and-commit strategy based on an initial activity map can lose access to informative electrodes as spontaneous activity redistributes over long-term recording. Per-recording turnover traces, firing-rate map correlations, and activity-centroid drift are shown in supplementary figure~\ref{supp-fig:nonstationarity}.

\subsection{Adaptive selection improved long-term activity capture}

We next asked whether adaptive electrode selection could recover more spontaneous activity than fixed or heuristic channel assignment under the same recording budget (figure~\ref{fig3}(C,D)). In the downscaled offline replay, each method selected 100 electrodes from 529 candidate electrodes at every recording step. Performance was measured by the attainment score, which compares the activity captured by the selected electrodes with the activity captured by the oracle subset for the same time window.

After an initial exploratory phase, the Bayesian adaptive method captured more activity than the baseline strategies and maintained higher oracle-relative attainment during the later recording period (figure~\ref{fig3}(C)). In oracle-relative terms, it achieved the highest attainment among the tested strategies and consistently outperformed static selection, random selection, and combinatorial $\varepsilon$-greedy selection. The improvement over static selection is consistent with the non-stationary structure of the recordings: electrodes that were active early in the recording did not necessarily remain the most informative throughout the full 34-hour period.

At the final 34 h time point, the proposed method exceeded static selection by 17.2 percentage points in attainment across the nine recordings (two-sided exact paired sign-flip randomization test, $p=0.004$; figure~\ref{fig3}(D)). The method consistently exceeded the baseline strategies, suggesting that the benefit was not limited to a single recording or activity pattern. This robustness is important for long-term HD-MEA experiments, where the spatial distribution of spontaneous activity can change gradually or abruptly. Per-recording replay traces are shown in supplementary figure~\ref{supp-fig:per_recording_replay}, and ablation and parameter-sensitivity analyses are shown in supplementary figures~\ref{supp-fig:ablation} and \ref{supp-fig:hyperparameter_sensitivity}.

Random selection provided a reference for unguided sampling, while static selection represented the common scan-and-commit strategy. Combinatorial $\varepsilon$-greedy selection improved on purely static assignment by allowing occasional exploration, but it did not match the Bayesian method. Thompson sampling increased exploration when posterior uncertainty was large and reduced it when the evidence for high activity was more stable, allowing the selected electrode set to follow changes without abandoning high-yield electrodes unnecessarily.

\subsection{Representative online recording of the first network burst}

We next tested whether the proposed adaptive policy could be executed online during an ongoing HD-MEA recording (figure~\ref{fig4}). Recording began 30~h after cell seeding, and the routed electrode configuration was updated every 30~min using the Bayesian hyperparameters that performed best in the offline replay; the online channel budget was 1,024 electrodes (table~\ref{tab:algorithm_parameters}). At recording onset, the selected electrodes showed no clear network-level activity. Activity then emerged gradually: sporadic and spatially restricted events appeared first, followed by broader population events and synchronized network bursts.

This progression is consistent with previous MEA studies in which \citet{chiappalone2006dissociated} reported spontaneously correlated activity during in vitro development and \citet{wagenaar2006extremely} described rich bursting patterns in cortical cultures. We therefore focused on the first synchronized burst after recording onset because it marks the transition from sparse local activity to coordinated network activity. In this online recording, the first burst was visible as a population-level increase in the electrodes selected by the proposed method, illustrating that the policy could be deployed in real time and preserve an early network-wide event.

\subsection{Activity centroids summarized burst trajectories}

To summarize the spatial structure of the online recording, we computed activity-centroid trajectories for the detected burst intervals in figure~\ref{fig4}. This analysis follows the idea of the center of activity trajectory, which has been used to reduce MEA spatiotemporal activity patterns to location-weighted trajectories \cite{chao2007region}. Spike event frame numbers were converted to time in milliseconds using the 20~kHz sampling rate, and each event was assigned the physical position of its electrode channel from the recording metadata. Each burst interval was then divided into 5~ms bins; bins containing fewer than 10 spikes were excluded from the trajectory calculation.

For each retained bin $b$, the activity centroid was computed as the mean position of all spikes in that bin,
\begin{equation}
\mathbf{x}_{\mathrm{act}}(b)
=
\frac{1}{|B_b|}
\sum_{j \in B_b}
\mathbf{x}_{c_j},
\end{equation}
where $B_b$ is the set of spikes in bin $b$, $c_j$ is the electrode channel of spike $j$, and $\mathbf{x}_{c_j}$ is the two-dimensional electrode position. Because the average is taken over spikes rather than unique electrodes, repeated spikes on the same electrode contribute repeatedly. The centroid is therefore equivalent to a spike-count-weighted center of activity, with
$x_{\mathrm{act}}$ and $y_{\mathrm{act}}$ given by the mean of the spike positions along the two array axes.

For each burst, the time-ordered centroid sequence formed a spatial trajectory. For visualization, trajectories were smoothed with a polynomial fit of degree at most three. The centroid trajectory complemented the firing-rate analysis: firing-rate traces indicate how much activity was captured, whereas the activity centroid indicates where that activity was expressed on the array. In the representative recording, centroid trajectories were sparse and variable around the first burst, but gradually became more recurrent and followed similar paths in later windows after burst emergence (figure~\ref{fig4}). In this representative recording, this representation reduced each burst to a spatial trajectory that could be followed over many hours. Optional supplementary movies 1--3 provide time-lapse visualizations of the selected-electrode map and representative activity-centroid trajectories (supplementary table~\ref{supp-tab:supplementary_movies}).
\end{multicols}

\section{Discussion}

\begin{multicols}{2}
\subsection{Adaptive sensing of non-stationary activity}

This study treated HD-MEA electrode allocation as an adaptive sensing problem for long-term recordings of spontaneous activity. The offline replay provides the main quantitative validation, whereas the online recording demonstrates feasibility of adaptive routing in a representative preparation. In offline replay of densely routed recordings, the results show that a Bayesian selection strategy can capture a large fraction of the activity available to an oracle while using only a limited number of recording channels. The main advantage is that the method does not assume a fixed set of informative electrodes; the configuration is updated as new observations arrive.

This distinction is important for dissociated neuronal cultures because spontaneous activity is structured but not stationary. Bursts may recur, but their spatial expression can vary across time, and gradual changes in culture state can shift the electrodes that provide the most useful signals. This interpretation is consistent with long-term HD-MEA observations that mature dissociated cultures can show chronic co-variation between network configuration and activity, suggesting that electrode usefulness may change with slow network reorganization rather than only with short-term firing-rate fluctuations \cite{okawa2015chronic}. A scan-and-commit strategy can work well when activity remains stable, but it has no mechanism for recovery once the informative region moves away from the initially selected electrodes. By maintaining uncertainty over electrode activity and discounting older observations, the proposed method can continue to test alternatives while preserving high-yield sites.

The comparison with combinatorial $\varepsilon$-greedy selection also suggests that exploration alone is not sufficient. Random exploration can discover changes, but it does not use uncertainty in a targeted way. Thompson sampling links exploration to the posterior distribution: electrodes are more likely to be sampled when they are either estimated to be active or remain uncertain enough to be plausibly useful. This property is well matched to long-term HD-MEA recordings, where the cost of each channel assignment is high and the activity landscape changes over time.

A recent-rate ranking strategy would be a strong baseline if recent spike counts were available for all electrodes. In the actual HD-MEA setting, however, activity on unselected electrodes is not observed. Obtaining an updated full-array firing-rate map would require a new scan of the array; in our setup, covering the 26,400-electrode array required 26 configurations of 1,024 electrodes recorded for 60 s each, corresponding to approximately 26 min per scan. Thus, frequent rescanning would consume a substantial fraction of the recording time and could miss transient population events during non-targeted scan configurations. We therefore focused on policies that update electrode allocation from partial observations under the same fixed channel budget, rather than assuming free access to recent full-array activity.

\subsection{Spike count as an initial selection objective}

The activity-centroid analysis provides another way to interpret why maximizing captured events is useful. Let $\mathbf{x}$ denote the position of a spike event, and let $\mathbf{m}$ be the centroid computed from all events in a time window. If the selected electrodes capture a fraction $p_s$ of events, with centroid $\mathbf{m}_s$, and the missed events have centroid $\mathbf{m}_c$, then the full centroid can be written as
\begin{equation}
\mathbf{m}=p_s\mathbf{m}_s+(1-p_s)\mathbf{m}_c .
\end{equation}
Therefore, the error of the centroid estimated from selected electrodes alone satisfies
\begin{equation}
\|\mathbf{m}_s-\mathbf{m}\|
=
\frac{1-p_s}{p_s}
\|\mathbf{m}-\mathbf{m}_c\| .
\end{equation}
This relation shows that, for bounded spatial activity, increasing the fraction of captured events directly reduces the possible centroid bias. Thus, a selection objective based on captured spike count is not only a measure of recording yield; it is also aligned with preserving center-of-activity trajectories under a limited channel budget. We used spike count as the first objective because it is directly observable, hardware-independent, and does not require assumptions about unit sorting or downstream event labels. The argument does not remove the need for spatially aware objectives in future work, but it explains why high event capture is a reasonable initial objective for long-term centroid-based analysis.

At the same time, spike-count maximization can over-sample highly active regions and under-sample lower-rate sites that are spatially or temporally informative. Depending on the experiment, future selection objectives could combine spike count with spatial diversity, sensitivity to burst onset, mutual information, propagation reconstruction error, uncertainty reduction, or novelty and change-point detection. A post hoc spatially regularized analysis is provided in supplementary figure~\ref{supp-fig:spatial_regularization}. These alternatives would allow the same adaptive framework to target burst detection, propagation tracking, unit isolation quality, or information gain rather than recording yield alone.

\subsection{Limitations and future directions}

Several limitations should be noted. First, the main quantitative evaluation used densely routed recordings to simulate a reduced channel budget offline. This was necessary for oracle-based comparison. The representative online run demonstrated online adaptive updates in one recording; however, systematic full-array online evaluation across multiple recordings remains future work. Scaling online implementation will require accounting for hardware switching latency, routing constraints, and the timing at which spike counts become available. Second, the current model treats electrodes independently. Incorporating spatial correlations could improve performance when activity propagates across neighboring electrodes or when neighboring electrodes provide redundant information.

Although this study focused on in vitro HD-MEAs, the same adaptive-sensing formulation may also be relevant to high-density in vivo probes, including Neuropixels-style devices, where dense recording sites can exceed the number of simultaneously usable channels and channel selection may need to adapt to unit yield, signal quality, or behavioral relevance \cite{jun2017fully,choi2020optimal}. In that setting, the reward would not need to be the total spike count; it could be defined from unit isolation quality, spike-sorting stability, task information, or information gain. Such extensions would require validation under in vivo constraints such as probe drift, tissue motion, and changing behavioral state, but they follow the same principle of choosing the next measurement from partial observations under a fixed channel budget.

Even with these limitations, the proposed framework provides a practical route toward adaptive HD-MEA recording. Rather than using dense electrode arrays only for an initial scan, the array can be treated as a resource that is revisited throughout the experiment. This view is useful for long-term culture recordings, where the target is not a static map of active sites but the evolving dynamics of the network.
\end{multicols}

\section{Conclusion}

\begin{multicols}{2}
We developed a Bayesian adaptive electrode-selection framework for long-term spontaneous HD-MEA recordings from dissociated neuronal cultures. By formulating electrode allocation as a sequential subset-selection problem, the method updates electrode activity estimates from observed spike counts and reallocates the fixed channel budget over time. This formulation directly addresses a practical limitation of HD-MEA experiments: the array provides dense spatial coverage, but only a subset of electrodes can be routed for simultaneous recording.

Using densely routed recordings as offline reference data, we first confirmed that the active electrode set was not stable: across nine recordings, top-100 active-electrode turnover reached 47.8\% at 34 h. The proposed method then captured a larger fraction of oracle activity than static, random, and combinatorial $\varepsilon$-greedy strategies, and exceeded static selection by 17.2 percentage points at the final 34 h time point. These results indicate that uncertainty-aware exploration and temporal discounting are useful for following non-stationary activity patterns over many hours in replayed recordings.

Representative online analysis further illustrated that adaptive selection could preserve biologically relevant transient events, including the first synchronized burst in one recording, and could provide activity maps suitable for downstream spatial summaries such as center-of-activity trajectories. Thus, the benefit of adaptive selection is not limited to increasing total spike yield; it may also help preserve temporal and spatial readouts that are important for interpreting long-term changes in cultured neuronal networks.

Taken together, these findings suggest that spontaneous activity in dissociated neuronal networks should not be treated as a static spatial target. Informative electrodes changed over hours, and adaptive allocation allowed the recorded subset to follow this evolving activity more effectively than scan-and-commit strategies. Thus, the present study provides a proof of concept for adaptive sensing of evolving neural activity under hardware readout constraints. Future work should extend the framework to systematic full-array real-time evaluation across multiple recordings, incorporate spatial correlations between neighboring electrodes, and define selection objectives tailored to specific experimental goals such as burst initiation, propagation analysis, or information-maximizing recording.
\end{multicols}

\section*{Acknowledgments}

This work was supported by JSPS KAKENHI (24K20854, 25H02600, 26H02517), AMED (24wm0625401h0001), the Asahi Glass Foundation, and the Secom Science and Technology Foundation.

\section*{Author Contributions}

Conceptualization: KT, DA, HT. Methodology: KT, DA. Software: KT. Validation: KT, DA. Formal analysis: KT, DA. Investigation: KT, DA. Resources: DA, HT. Data Curation: KT, DA, HT. Writing - Original Draft: KT. Writing - Review \& Editing: DA, HT. Visualization: KT, DA. Supervision: DA, HT. Project administration: HT. Funding acquisition: KT, DA, HT.

\section*{Data Availability}

Core analysis code implementing the proposed Bayesian electrode selection framework is available at \url{https://github.com/nelab-utokyo/dynamic_sampling}. The repository will be updated with the final analysis scripts upon publication of the peer-reviewed article. Code interfacing with proprietary hardware APIs (MaxOne, Maxwell Biosystems) is available from the corresponding author upon reasonable request. The data that support the findings of this study will be made openly available upon publication of the peer-reviewed article.

\section*{Conflict of Interest Statement}

The authors declare no conflict of interest.

\section*{ORCID iDs}

\begin{flushleft}
Kazushi Takehana \orcidicon{0009-0000-8487-3826} \url{https://orcid.org/0009-0000-8487-3826} \\
Dai Akita \orcidicon{0009-0009-6445-2587} \url{https://orcid.org/0009-0009-6445-2587} \\
Hirokazu Takahashi \orcidicon{0000-0002-1834-3832} \url{https://orcid.org/0000-0002-1834-3832}
\end{flushleft}

\section*{Supplementary Material}

Supplementary material is included with this arXiv version as a separate file. File title: Supplementary material. File description: Recording metadata, candidate-region maps, spike and burst detection details, offline replay definitions, non-stationarity analyses, per-recording replay results, ablation analyses, hyperparameter sensitivity, spatially regularized objective analyses, and optional supplementary movie descriptions for adaptive HD-MEA electrode selection.

\bibliographystyle{unsrtnat}
\bibliography{references}

\end{document}